\documentclass[aps,pre,preprint,superscriptaddress,showpacs,floatfix]{revtex4}
\usepackage{amssymb}
\usepackage{epsfig}
\usepackage{amsmath}

\begin{document}

\title{ Effects of gradient coupling on amplitude death in
nonidentical oscillators}

\author{Weiqing Liu}
\affiliation{School of Science, Jiangxi University of Science and
Technology, Ganzhou 341000, China}
\author{Jinghua Xiao}
\affiliation{School of Science, Beijing University of Posts and
Telecommunications, Beijing 100876, China}
\author{Lixiang Li}
\affiliation{Key Laboratory of Network and Information Attack and
Defence Technology of MOE, Beijing University of Posts and
Telecommunications, Beijing 100876, China}
\author{Ye Wu}
\affiliation{School of Science, Beijing University of Posts and
Telecommunications, Beijing 100876, China}
\author{Min Lu}
\affiliation{School of Science, Jiangxi University of Science and
Technology, Ganzhou 341000, China}
\date{\today}

\begin{abstract}
 In this work, we investigate gradient coupling effect on amplitude death in an array of
 $N$ coupled nonidentical oscillators with no-flux boundary conditions
and periodic boundary conditions respectively. We find that the
effects of gradient coupling on amplitude death in diffusive coupled
nonidentical oscillators is quite different between those two
boundaries conditions. With no-flux boundary conditions, there is a
system size related critical gradient coupling $r_c$ within which
the gradient coupling tends to monotonically enlarge the amplitude
death domain in the parameter space. With the periodical boundary
conditions, there is an optimal gradient coupling constant $r_o$ to
realize largest AD domain. The gradient coupling first enlarges then
decreases the amplitude death domain of diffusive coupled
oscillators. The amplitude death domain of parameter space are
analytically predicted for small number of gradient coupled
oscillators.
\end{abstract}


\maketitle

\newpage
\textbf{I. Introduction}

The model of coupled nonlinear oscillators provides a simple but
powerful paradigm for understanding of collective behaviors such as
emergent behavior which are widely explored in the interacting large
number of natural oscillators. Therefore, the study of coupled
nonlinear oscillators has become a very hot topic in nonlinear
sciences and many other interdisciplinary fields such as physical,
chemical, biological, and even social sciences\cite{win,kur,pik}.

Ensembles consisted of different types of coupled oscillators
exhibit various of collective behaviors. Among, the amplitude
death(AD), which refers to a situation where individual oscillators
cease oscillating when coupled and go to an equilibrium solution
instead, has been actively investigated since the appearance of
Ref\cite{bar}. AD plays a crucial role in a lot of real systems, for
instance, it has been extensively found in synthetic genetic
networks\cite{ull,kos1,kos2}, where the AD implies a constant
protein expression and the system multi-stability with AD is
believed to improve the adaptability and robustness of the cellular
population. Significant progress was achieved in theoretical and
numerical analysis of AD in systems of oscillators with various
coupling schemes such as all-to-all coupling\cite{erm},diffusively
coupling\cite{yang,rub,atay} even in the complex networks where the
effects of topological properties of the network on partial AD
dynamics are explored\cite{hou,liu}. AD can be eliminated by
introducing random links and the influence of spatial disorder on AD
in oscillator arrays with local couplings was extended. The
desynchronization-induced AD are weakened considerably by
introducing the random deviation from a linear trend of frequencies
in array of diffusively coupled limit-cycle oscillators with a
regular monotonic trend of natural frequencies\cite{rub}.

The gradient coupling, one of anisotropic coupling, is of practical
importance in many situations, such as in hydrodynamics flows with
sloping channels and in plasma systems with electromagnetic fields.
The gradient coupling has effects in the control of spatiotemporal
chaotic systems \cite{xiao1,xiao2} and in the synchronization of
coupled nonlinear oscillators \cite{zhan1,zhan2,zhan3}. It is
observed that as the gradient coupling constant increases, the
networks' synchronizability is possibly enhanced or decreased or
even a optimal value of coupling constant exists for best
synchronizability \cite{mot,xin,xin1}. Therefore, it is also
significant to explore the effects of gradient coupling on another
collective behavior, the AD dynamics, in array of diffusively
coupled oscillators. However, most of previous studies on AD have
been confined only to the cases of coupled oscillators with a
homogeneous identical coupling from their neighbors, and the effect
of anisotropic coupling has seldom been studied until in the recent
published article\cite{zou}, the effects of gradient coupling on the
time-delay induced AD is explored. The gradient coupling tends to
monotonically reduce the domain of delay-induced death island.
Typically for the occurrence of AD, one of the following two
conditions is needed: the time-delayed coupling \cite{ram}or the
parameter mismatches\cite{hou,yang,liu}. In addition to these two
general conditions, recent studies revealed that AD may also happen
by dynamic coupling \cite{kon} or conjugate coupling
\cite{zou1,kar}. Since frequency mismatches is widely existing in
the natural world. It is meaningful to ask how does the gradient
coupling influence the frequency mismatches induced AD in the array
of diffusively couple oscillators. Is the anisotropic coupling
beneficial to decrease the minimal frequency mismatch needed for AD
for given diffusive coupling?  To answer those question, the
influences of gradient coupling on the AD domain of oscillators with
frequency mismatches are explored. We find that the effects of
gradient coupling on AD in diffusive coupled oscillators is strongly
related to the boundary conditions which has effects on the
synchronization ability of the diffusively coupled oscillators as
discussed in Ref.\cite{wen}. With no-flux boundary conditions, the
gradient coupling constant monotonically enlarge the AD domain until
it is larger than the system size related critical value $r_{c}$.
with periodic boundary conditions, the increasing gradient coupling
first enlarge then decrease the domain of AD. There is an optimal
system size related gradient coupling constant $r_o$ to realize
largest AD domain. The parameter space of AD are analytically
predicted for small number of gradient coupled oscillators.

\textbf{II. Gradient coupling model}

Consider the general form of N gradient coupled nonidentical
oscillators with
\begin{eqnarray}\label{eq1}
\dot{X_j} &=&f(\omega_{j},X_j)+(\epsilon+r)(X_{j+1}-X_{j})+
(\epsilon-r)(X_{j-1}-X_{j})
\end{eqnarray}
for $j=1, . . . , N$($N\ge3$), where $X_{j}$ represents the state
vector of the $jth$ element; $\epsilon$ and r are the diffusive and
gradient coupling strengths respectively. The uncoupled units
$\dot{X_j}=f(\omega_{j},X_j)$ have non-stationary behaviors and
meanwhile, accompanying an unstable focus $X^{*}$. $\omega_{j}$ is
the natural frequencies of uncoupled oscillators.  With the
frequency mismatches, the AD may occur in certain range of coupling
constant region when the coupling interaction turns the formerly
unstable focus $X^{*}$ stable. The distribution of the frequencies
has great effect on the AD dynamics. i.e. Ref. \cite{rub}, the AD in
the diffusively coupled oscillators with a regular monotonic trend
of natural frequencies can be weakened by the disorder of
frequencies distribution. The transition process from partial AD to
complete AD  in diffusively coupled oscillators with frequency
mismatches are explored in Ref. \cite{yang}. For the random
distribution frequencies, the frequency deviation has greatly
influenced the AD behaviors. Without losing generality, we consider
$N$ coupled nonidentical Landau-Stuart oscillators with the no-flux
boundary condition $z_{N+1}=z_{N}, z_{0}=z_{1}$ and the periodical
boundary condition $z_{N+1}=z_{1}, z_{0}=z_{N}$ respectively. The
coupled system is presented as follows.
\begin{eqnarray}\label{eq2}
\dot{z_j} &=&(1+i\omega_{j}+|z_j|^2)z_j+(\epsilon+r)(z_{j+1}-z_{j})+
(\epsilon-r)(z_{j-1}-z_{j}),j=1,...,N
\end{eqnarray}
where $i$ is the imaginary, $z_j$ is the complex variables, $w_j$
are the intrinsic frequencies of single uncoupled oscillators. For
simplicity, we suppose the coupled oscillators have a regular
monotonic trend of natural frequencies $w_j=w_{1}+(j-1)\delta\omega,
j=1,2,...,N, w_1=1$,$\delta\omega>0$. In the absence of coupling,
each oscillator has an unstable focus at the origin $|z_j|=0$ and an
attracting limit cycle $z_j(t)=e^{i\omega_jt}=x(t)+iy(t)$ with
different oscillating frequencies $\omega_j$.

  For $r=0$, Eq.\ref{eq2} is diffusive coupled system as discussed in Ref.
  \cite{rub,yang}, where the necessary condition of AD is
  $\epsilon>0.5$. To explore the influence of gradient coupling on
  the domain of AD, the stability of the complete AD can be analyzed
by linearizing the Eq. \ref{eq2} at $|z_j|=0,j=1,2,...,N$.
\begin{eqnarray}\label{eq3}
\dot{\eta}_j(t)=(1-2\epsilon+i\omega_j)\eta_j(t)+(\epsilon+r)\eta_{j+1}(t)+(\epsilon-r)\eta_{j-1}(t)
\end{eqnarray}
With denotations of $\eta(t)=(\eta_1(t),\eta_2(t), . . .
,\eta_n(t))'$ , the above equations can be rewritten as follows.
\begin{eqnarray}\label{eq4}
\dot{\eta}(t)=B\eta(t)
\end{eqnarray}
where for non-flux boundary condition

$B=B_n=\left(\begin{array}{ccccc}1-2\epsilon+i\omega_1 &
\epsilon+r&& \\
\epsilon-r &1-2\epsilon+i\omega_2&\epsilon+r,&
\\&\epsilon-r&1-2\epsilon+i\omega_3&\epsilon+r\\ &...&...&...& \\
&&\epsilon-r&1-2\epsilon+i\omega_N&
\end{array}\right)$.\\

or for periodical boundary condition

$B=B_p=\left(\begin{array}{ccccc}1-2\epsilon+i\omega_1 & \epsilon+r&&\epsilon-r \\
\epsilon-r &1-2\epsilon+i\omega_2&\epsilon+r&
\\&\epsilon-r&1-2\epsilon+i\omega_3&\epsilon+r\\ &...&...&...& \\
\epsilon+r&&\epsilon-r&1-2\epsilon+i\omega_N&
\end{array}\right)$

Assume that B can be diagonalized by the matrix P:
\begin{eqnarray}\label{eq5}
 P^{-1}BP=diag(\lambda_0,\lambda_1,...,\lambda_{N-1})
\end{eqnarray}
 where $\lambda_k, k=0,1, . . . ,N-1
 $ are the eigenvalues of the matrix B. The necessary condition for
 AD of Eq.\ref{eq2} is that all real part of the eigenvalues $Re(\lambda_k)<0, k=0,1, . . .
 ,N-1$. Therefore, the region of AD state is completely determined
 by the critical line of all $Re(\lambda_k)<0, k=0,1, . . . ,N-1$.
 However, it is difficult to diagonalize the matrix B analytically when $N$ is large.
 To investigate the effects of gradient coupling on AD domain, we mainly resort to numerical simulations for large
 $N$. Moreover, the AD domain can be analytically presented for small
 system size $N$.

\textbf{III. The non-flux boundary condition}

\begin{figure}
\includegraphics[width=14cm]{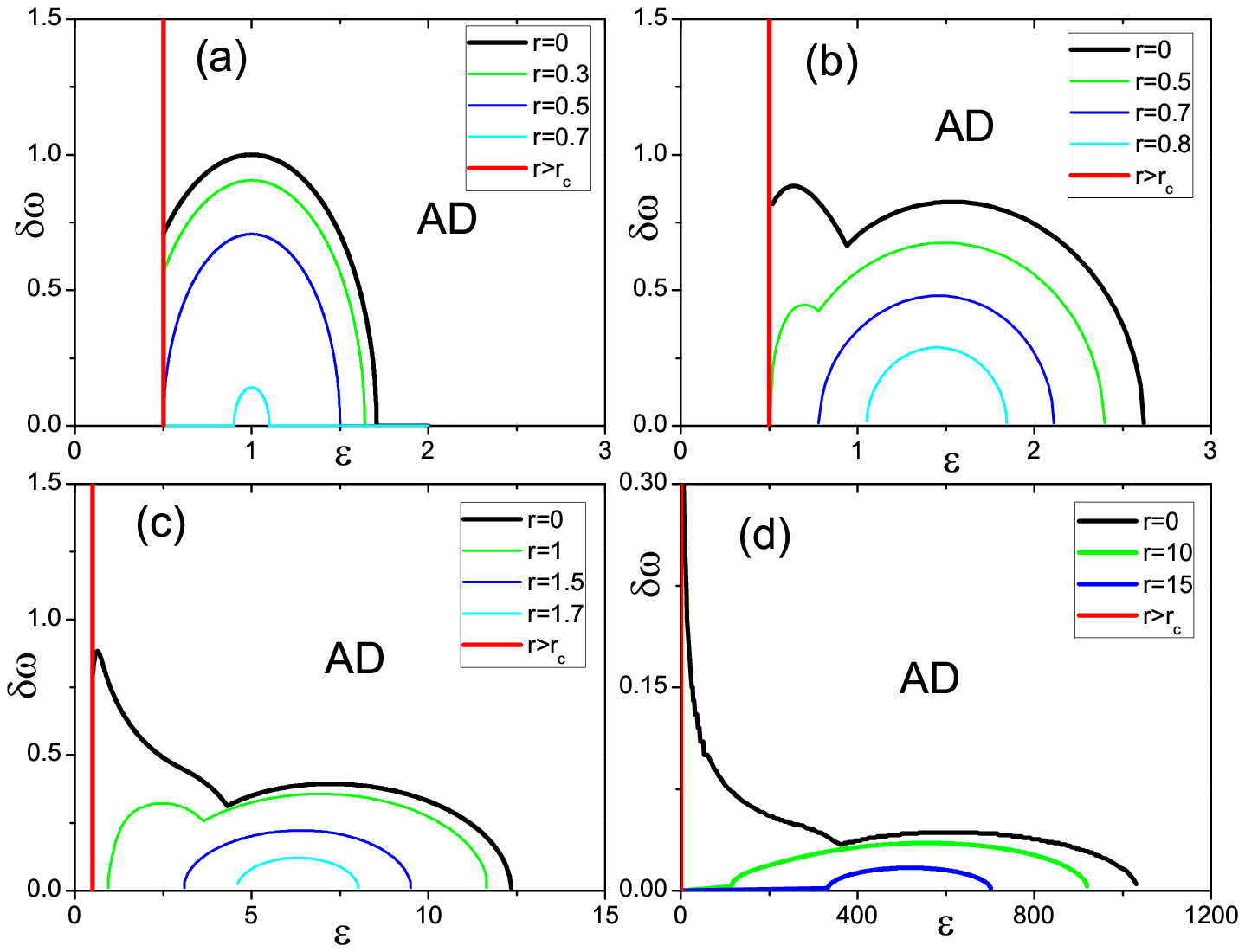}
\caption{(Color online) The critical line of the AD in coupled
system with non-flux boundary condition. The AD domain is the right
and up part of the area enclosed by the line
$\epsilon>0.5,\delta\omega>0$ and the critical lines of
corresponding $r$. (a) $N=3$,$r=0$(black line),r=0.3 (green
line),r=0.5 (blue line),r=0.7 (cyan line); (b) $N=4$,$r=0$ (black
line),r=0.5 (green line),r=0.7(blue line),r=0.8 (cyan line);
(c)$N=10$, $r=0$ (black line),r=1 (green line),r=1.5 (blue
line),r=1.7 (cyan line); (d)$N=100$,$r=0$ (black line),r=1 (green
line),r=2 (blue line),r=3 (cyan line),r=5 (magenta line).}
\label{fig_1}
\end{figure}

\begin{figure}
\includegraphics[width=14cm]{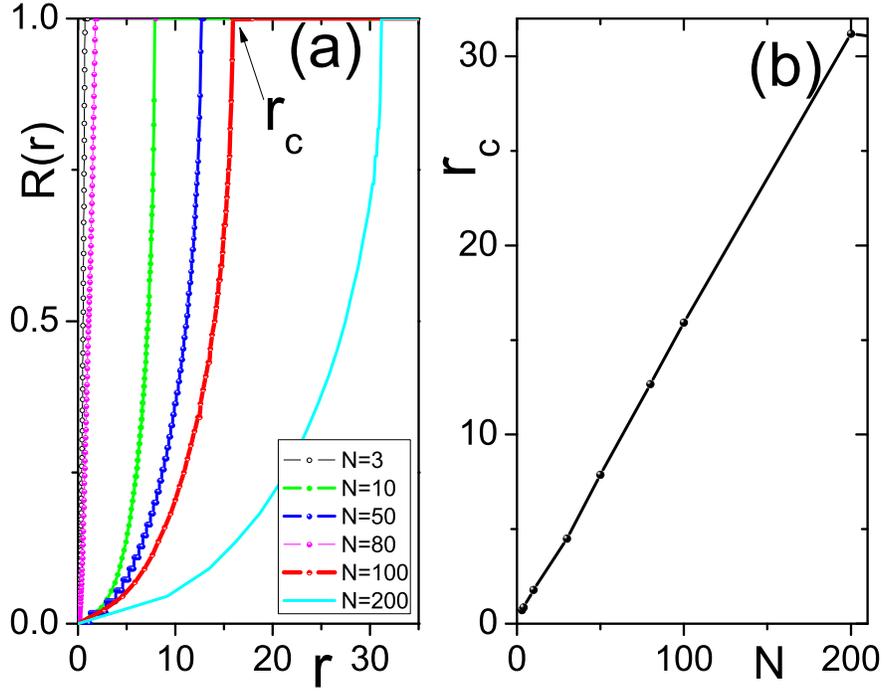}
\caption{(Color online) (a) The normalized scaling factor $R(r)$ vs
r for $N=3, 10, 50, 80,100,200$, respectively. $R(r)=1-S(r) /S(0)$
with $S(r)$ standing for the area of non-AD island in parameter
space $\epsilon \sim \delta \omega$ for $\epsilon>0.5$. For each
given system size N, a monotonic increasing of R(r) vs r and a
critical value $r_c$ are clear. All the data are numerically
obtained by directly integrating the coupled system Eq.\ref{eq2}.
(b) The critical gradient coupling constant $r_c$ vs system size N.
$r_c$ is linearly increasing with N.} \label{fig_2}
\end{figure}

 Let's firstly consider the gradient coupling effects on AD in the coupled system with non-flux boundary condition.
The AD domain of parameter space $\epsilon \sim \delta \omega$ are
calculated for different $N$ , $r$ in Eq.\ref{eq2} as shown in
Fig.\ref{fig_1}(a)$\sim$(d) for $(N=3,4,10,100)$ respectively. The
AD domain is the right and above part of area enclosed by
$\epsilon\ge0.5$(red solid line) and the curve line of corresponding
$r$. The area enclosed by $\epsilon>0.5$ and the black solid curve
for $r=0$ (marked with AD) in Fig.\ref{fig_1} are the AD domain for
$r=0$ which is right the AD domain of diffusive coupled system.
Interestingly, we find that the AD region is monotonically expanding
as the gradient coupling $r$ is gradually increasing until $r=r_c$,
where $r_c$ is related to the systems size $N$. Thus, for an
arbitrary given system size $N$, the gradient coupling may
monotonically increase the AD domain for $r<r_c$. However, when
$r\ge r_c$, the AD domain becomes the area enclosed by all
$\epsilon>0.5$ and $\delta \omega>0$ and keeps constant for
increasing gradient coupling strength $r$. Therefore, the gradient
coupling tends to minimize the frequency mismatch needed for AD in
diffusively coupled oscillators. According to Fig.\ref{fig_1}, there
is another remarkable thing that smaller frequency mismatches
$\delta \omega$ and larger diffusive coupling strength is needed for
AD in the gradient coupled system with larger system size $N$ for
arbitrary given constant $r$. However, based on these observations,
we may predict the critical gradient coupling $r_c$ for each $N$. A
normalized scaling factor is defined as $R(r)=1-S(r) /S(0)$, where
$S(r)$ denotes the area of non-amplitude death island in the
parameter space $\epsilon \sim \delta \omega $ for $\epsilon>0.5$.
Obviously, $R(0)=0$. The relationship between $R(r)$ and r are
presented in Fig.\ref{fig_2}(a) for different system size $N$.
$R(r)$ monotonically increases with increasing $r$ until $r=r_c$,
$R(r_c)=1$. The critical gradient coupling $r_c$ has positive linear
relationship with the system size $N$ as shown in
Fig.\ref{fig_2}(b).

\begin{figure}
\includegraphics[width=14cm]{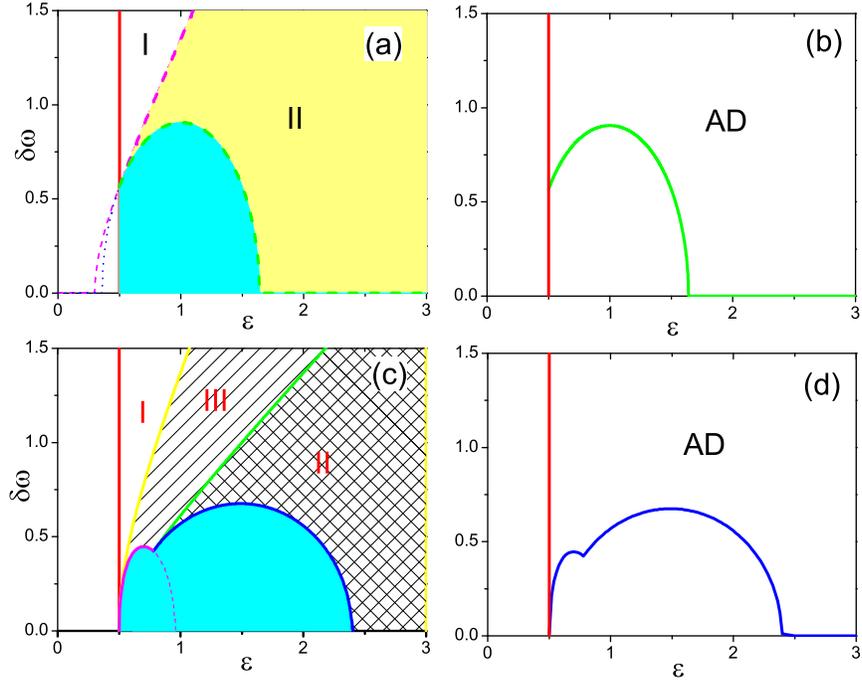}
\caption{(Color online) The boundary line of the AD domain in
coupled system with non-flux boundary condition. (a)Analytical
results of the AD domain for $N=3r=0.3$(area I and II). magenta
dashed line is Eq.\ref{eq7}, green dashed line is Eq.\ref{eq9}. (b)
Simulation results with parameters corresponding to (a). (c)
Analytical results of the AD domain for $ N=4,r=0.5$(area I,II and
III). (d) Simulation results with parameters corresponding to (c)}
\label{fig_3}
\end{figure}

 The AD domain of the gradient coupled system can be analytically
 presented according to $Re(\lambda_i)<0$(the eigenvalues of matrix $B_n$) when the size $N$ is small,
 for example, $N=3,4$. When $N=3$, the eigenvalues of matrix $B_n$ are presented as
 follows,
\begin{eqnarray}\label{eq6}
\lambda_1 & = & 1-2\epsilon+i(1+\delta \omega), \nonumber   \\
\lambda_{2,3}&=&
1-2\epsilon\pm\sqrt{2\epsilon^2-2r^2-\delta\omega^2}+i(1+\delta
\omega)
\end{eqnarray}
Let $Re(\lambda_i)<0$, the boundaries of the AD domain can be
determined by following equations.\\
(1). Area I, if
\begin{eqnarray}\label{eq7}
\delta\omega^2\ge2\epsilon^2-2r^2
\end{eqnarray}
then $\epsilon>0.5$ \\
(2). Area II, if
\begin{eqnarray}\label{eq8}
\delta\omega^2<2\epsilon^2-2r^2
\end{eqnarray} then
\begin{eqnarray}\label{eq9}
\epsilon & > & 0.5, \nonumber\\
\delta \omega^2 &  > & 1-2(\epsilon-1)^2-2r^2
\end{eqnarray}
Therefore, the AD domain for arbitrary given $r$ is composed of area
I and II(AD domain for $r=0.3$ is presented in Fig.\ref{fig_3}(a),
which is completely consistent with the numerical results as
presented in Fig.\ref{fig_3}(b)).  As $r$ increase from zero to
$r_c$, the AD area II enclosed by Eq.\ref{eq9} enlarges
monotonically.  As $r>r_c$, AD is stable for all $\epsilon>0.5$ and
$\delta\omega>0$ and keeps constant to increasing $r$. From
Eq.\ref{eq9}, the critical gradient coupling $r_c$ of $N=3$ can be
theoretically calculated as $\delta \omega^2\ge 0$ for all
$\epsilon>0.5$, that is, $1-2r^2=0$,($r_c=\sqrt{1/2}=0.707$).

 In the case of $N=4$, the eigenvalues of matrix $B_n$ can be given as
\begin{eqnarray}\label{eq10}
\lambda_{1,2,3,4} & = & 1-2\epsilon\pm0.5\sqrt{Q_{1,2}}+i(1+3\delta \omega/2),\nonumber\\
Q_{1,2} & = & 6\epsilon^2-6r^2-5\delta\omega^2\pm2\sqrt{P},\nonumber\\
P &=
&(\epsilon^2-r^2-2\delta\omega^2)(\epsilon^2-r^2-0.4\delta\omega^2)
\end{eqnarray}
The boundaries of the AD domain can also be presented as areas I,
II, and III according to $Re(\lambda_{1,2,3,4})\le0$.

(1) Area I,  enclosed by Eq.\ref{eq16};
\begin{eqnarray}\label{eq16}
\delta \omega ^2&\ge&(\epsilon^2-r^2)/0.4,\nonumber\\
\epsilon&>&0.5
\end{eqnarray}
(2) Area II, enclosed by Eq.\ref{eq17}.
\begin{eqnarray}\label{eq17}
\epsilon>0.5,\nonumber\\
\delta\omega ^2\le(\epsilon^2-r^2)/2,\nonumber\\
\delta\omega ^2>\frac{1}{9}(-20-6r^2+80\epsilon-74\epsilon^2\nonumber\\
+2\sqrt{1584\epsilon^2-512\epsilon-2240\epsilon^3+192\epsilon
r^2+64-48r^2+1189\epsilon^4-138\epsilon^2r^2-27r^4})
\end{eqnarray}
(3) Area III, enclosed by Eq.\ref{eq18}.\\
\begin{eqnarray}\label{eq18}
\epsilon&>&0.5,\nonumber\\
(\epsilon^2-r^2)/2&<&\delta\omega ^2<(\epsilon^2-r^2)/0.4,\nonumber\\
\delta\omega^2&>&\frac{1}{4}(-10-6r^2-34\epsilon^2+40\epsilon\nonumber\\
+2\sqrt{4r^4+16\epsilon^2r^2-24\epsilon
r^2+6r^2+124\epsilon^4-264\epsilon^3+210\epsilon^2-72\epsilon+9}
\end{eqnarray}

The AD domain for $r=0.5$ is marked as area I, II and III in
Fig.\ref{fig_3}(c) which is coincided with the numerical results in
Fig.\ref{fig_3}(d) well. According to Eq.\ref{eq17}, the critical
gradient coupling $r_c$ of $N=4$ can be determined by Eq.\ref{eq19}.

\begin{eqnarray}\label{eq19}
\frac{1}{9}(-20-6r^2+80\epsilon-74\epsilon^2 \nonumber\\
+2\sqrt{1584\epsilon^2-512\epsilon-2240\epsilon^3+192\epsilon
r^2+64-48r^2+1189\epsilon^4-138\epsilon^2r^2-27r^4})=0
\end{eqnarray}

The solution of Eq.\ref{eq19} is Eq.\ref{eq20}.
\begin{eqnarray}\label{eq20}
\epsilon=1\pm\frac{\sqrt{5}}{5}\pm\sqrt{0.7+0.3\sqrt{5}-(1+0.4\sqrt{5})r^2}
\end{eqnarray}
Since $\epsilon$ is real, then
$0.7+0.3\sqrt{5}-(1+0.4\sqrt{5})r^2\ge0$, that is
$r_c=\sqrt{\frac{0.7+0.3\sqrt{5}}{(1+0.4\sqrt{5}}}=0.8507$.

\textbf{IV. The periodical boundary condition}

\begin{figure}
\includegraphics[width=14cm]{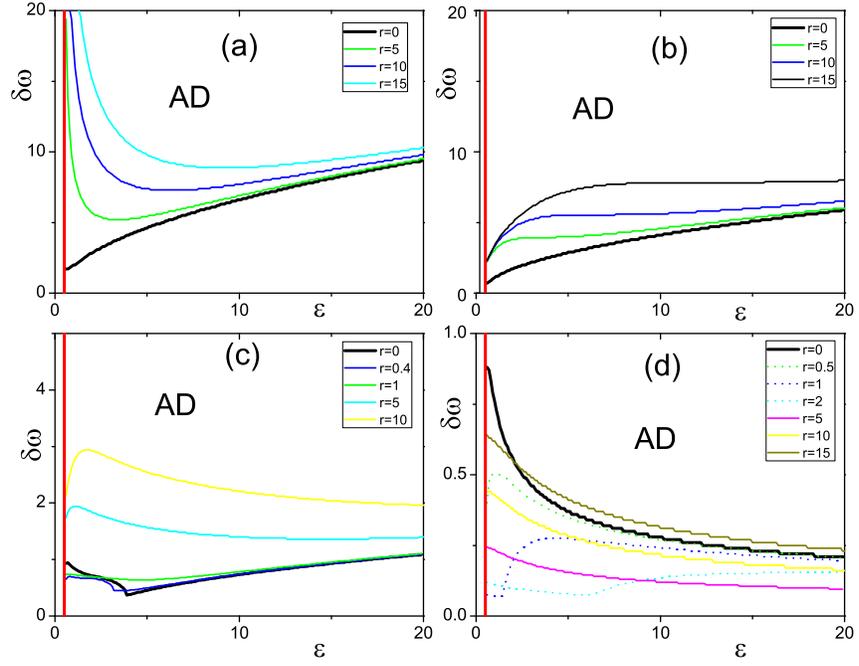}
\caption{(Color online) The critical line of the AD in coupled
system with periodical boundary condition. The AD domain is the
right and up part of the area enclosed by the line $\epsilon>0.5$
and the critical curve lines of corresponding $r$.(a) $r=0,5,10,15$
,$N=3$. (b) $r=0,5,10,15$, N=4. (c) $r=0,0.4,1,5,10$, N=10; (d)
$r=0,0.5,1,2,5,10,15$,N=100.} \label{fig_4}
\end{figure}

\begin{figure}
\includegraphics[width=14cm]{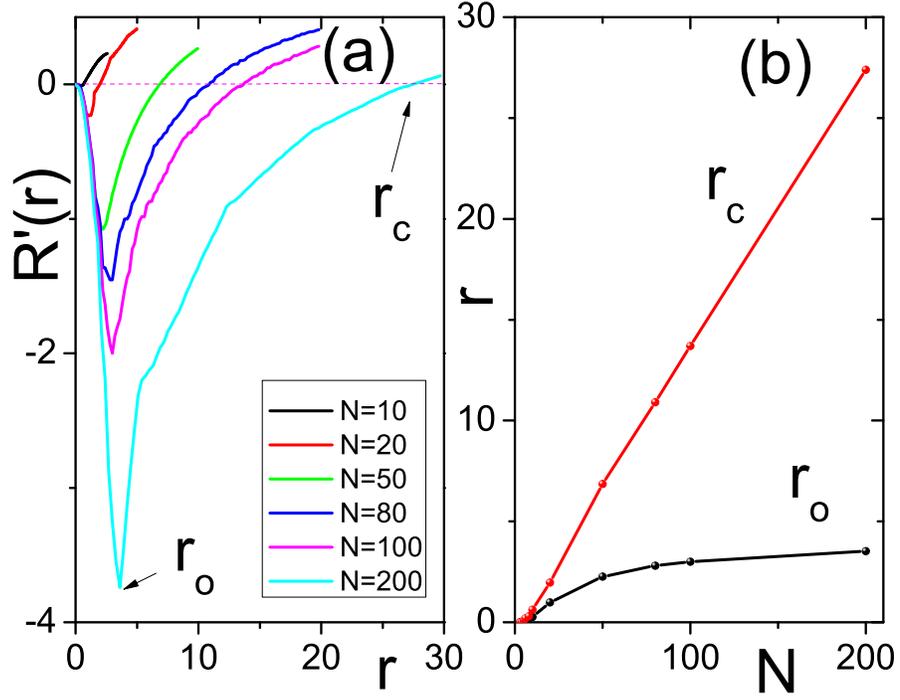}
\caption{(Color online)(a) The factor $R'(r)$ vs r for
$N=10,20,50,80,100,200$ respectively. $R'(r)=1-S(0) /S(r)$ with
$S(r)$ standing for the area of non-AD island in parameter space
$\epsilon \sim \delta \omega$ for $\epsilon
\in(0.5,20),\delta\omega\in(0,20)$. For each given system size
$N(N\ge5)$, $R'(r)$ firstly decrease to negative then increase to
positive. There is a optimal gradient coupling constant $r_o$ which
satisfy $R'(r_o)=R'_{min}$ and a critical gradient coupling constant
$r_c$ which satisfy $R'(r_c)=0$. (b) The optimal gradient coupling
constant $r_o$ and the critical gradient coupling constant $r_c$
versus system size $N$($r_o=0$ when $N=3,4)$.} \label{fig_5}
\end{figure}

Now let's explore the gradient coupled oscillators with periodical
boundary condition. By checking the AD domain of parameter space
$\epsilon \sim \delta \omega$ which is similar to that in Fig. \ref
{fig_1} for $N=3,4,10,100$ as shown in Fig.\ref{fig_4}(a)$\sim$(d),
where the AD region is the above part of area enclosed by line
$\epsilon>0.5$ and the curve line of corresponding $r$. One may find
that when $N=3,4$, the gradient coupling tends to decrease the AD
domain monotonically as shown in Fig.\ref{fig_4}(a)(b), while for
$N=10,100$, there is a critical gradient coupling constant $r_c$,
within which the gradient coupling first increase then decrease the
AD domain. That is to say, there is an optimal gradient coupling
constant $r_o$ with which the coupled system has the largest AD
domain in parameter space $\epsilon \sim \delta \omega$. However,
when $r>r_c$, the gradient coupling begin to shrink the AD domain of
formally diffusively coupled system.
 To quantified the critical gradient coupling constant $r_c$ and the optimal gradient coupling constant $r_o$
 for different size $N$, a factor is defined as $R'(r)=1-S(0)/S(r)$, where $S(r)$ is the area of the non-AD domain of corresponding
 $r$ for $\epsilon
\in(0.5,20),\delta\omega\in(0,20)$. (Obviously, $R'(0)=0$;
$\frac{dR'(r)}{dr}<0$ means the increasing $r$ enlarges AD domain;
When $R'(r)>R'(0)$, the gradient coupling have smaller AD domain
than the formally diffusively coupled system). Then $r_c$ is
determined by $R'(r_c)=0$ and the optimal gradient coupling constant
$r_o$ are determined by $\frac{dR'(r)}{dr}|_{r=r_o}=0$. The
relationship between $R'(r)$ and $r$ for different system size $N$
are presented in Fig.\ref{fig_5}(a). Obviously, there is system size
related $r_o$ and $r_c$. The detail relationship between $r_o$ and
$r_c$ and system $N$ are presented in Fig.\ref{fig_5}(b). $r_c$ is
linearly increase with $N$. $r_o=0$ for $N=3,4$ and increases
nonlinearly with increasing $N$.

Accordingly, the AD domain can be predicted for small system size
for example $N=4$. The eigenvalue of $B_p$ can be presented as
follows,
\begin{eqnarray}\label{eq21}
\lambda_{1,2,3,4}&=&1-2\epsilon\pm0.5\sqrt{T_{1,2}}+i(1+1.5\delta\omega), \nonumber \\
T_{1,2}&=&8\epsilon^2-8r^2-5\delta\omega^2\pm 4\sqrt{S}, \nonumber \\
S&=&\delta\omega^4-4\delta\omega^2(\epsilon^2-r^2)+4(\epsilon^2+r^2)^2.
\end{eqnarray}

The boundaries of the AD domain are determined by
$Re(\lambda_{1,2,3,4})<0$. The value of $Re(\lambda_{1,2,3,4})$ has
various forms for corresponding parameters.\\
(1) if $T<0$, i.e.
\begin{eqnarray}\label{eq22}
\delta\omega^2>\frac{4}{9}(-2r^2+2\epsilon^2+2\sqrt{r^4+34e^2r^2+e^4})
\end{eqnarray}
then $Re(\lambda_{1,2,3,4})=1-2\epsilon$\\
(2) if $T\ge0$, i.e.
\begin{eqnarray}\label{eq23}
\delta\omega^2\le\frac{4}{9}(-2r^2+2\epsilon^2+2\sqrt{r^4+34e^2r^2+e^4})
\end{eqnarray}
then
\begin{eqnarray}\label{eq24}
Re(\lambda_{1,2,3,4})&=&1-2\epsilon\pm0.5\sqrt{T_{1,2}}, \nonumber \\
T_{1,2}&=&8\epsilon^2-8r^2-5\delta\omega^2\pm
4\sqrt{\delta\omega^4-4\delta\omega^2(\epsilon^2-r^2)+4(\epsilon^2+r^2)^2}
\end{eqnarray}
Thus, the critical line of AD domain can be found by solving the
 equations $Re(\lambda_{1,2,3,4})\le0$. The AD domain consist of
two areas\\
(1) Area I  enclosed by solution of $\epsilon>0.5$ and Eq.\ref{eq22}; \\
(2) Area II enclosed by solution of Eq.\ref{eq23} and
Eq.\ref{eq25}($Re(\lambda_{1,2,3,4})\le0$ in Eq.\ref{eq24})
\begin{eqnarray}\label{eq25}
\delta\omega^2 & \ge & g(\epsilon,r),\nonumber \\
g(\epsilon,r)&=&\frac{2}{3}(-18\epsilon^2+20\epsilon-2r^2-5\nonumber \\
+2\sqrt{81\epsilon^4-144\epsilon^3+18\epsilon^2r^2+100\epsilon^2+16r^2\epsilon-32\epsilon+r^4-4r^2+4})
\end{eqnarray}

\begin{figure}
\includegraphics[width=14cm]{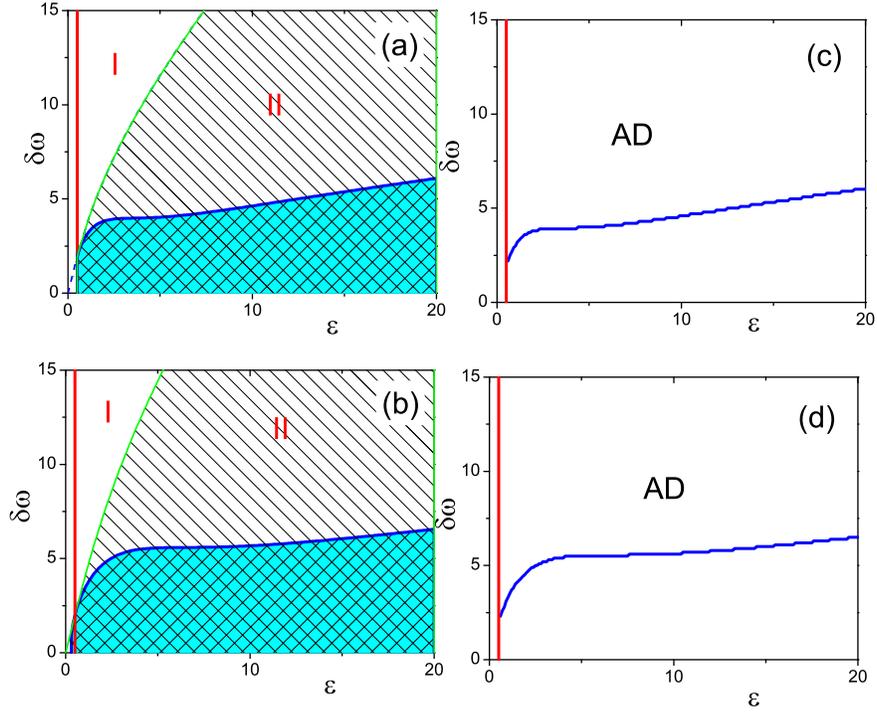}
\caption{(Color online) (a)(b)Analytical result of the critical line
of the AD domain in coupled system with periodical boundary
condition and $N=4$. $r=5$,$r=10$ respectively. The AD domain is
consisted of area I and area II. (c)(d) The simulation results of AD
domain with parameter corresponding to (a) (b) respectively. The
analytical results are consistent with the simulation results.}
\label{fig_6}
\end{figure}

The AD domain of $r=5,10$ is presented as area I and II in
Fig.\ref{fig_6}(a)(b), where area I is enclosed by $\epsilon>0.5$
and Eq.\ref{eq22}, while area II is enclosed by Eq.\ref{eq23} and
Eq.\ref{eq25}. It is also well consistent with the numerical results
as shown in Fig.\ref{fig_6}(c)(d) respectively.

\begin{figure}
\includegraphics[width=14cm]{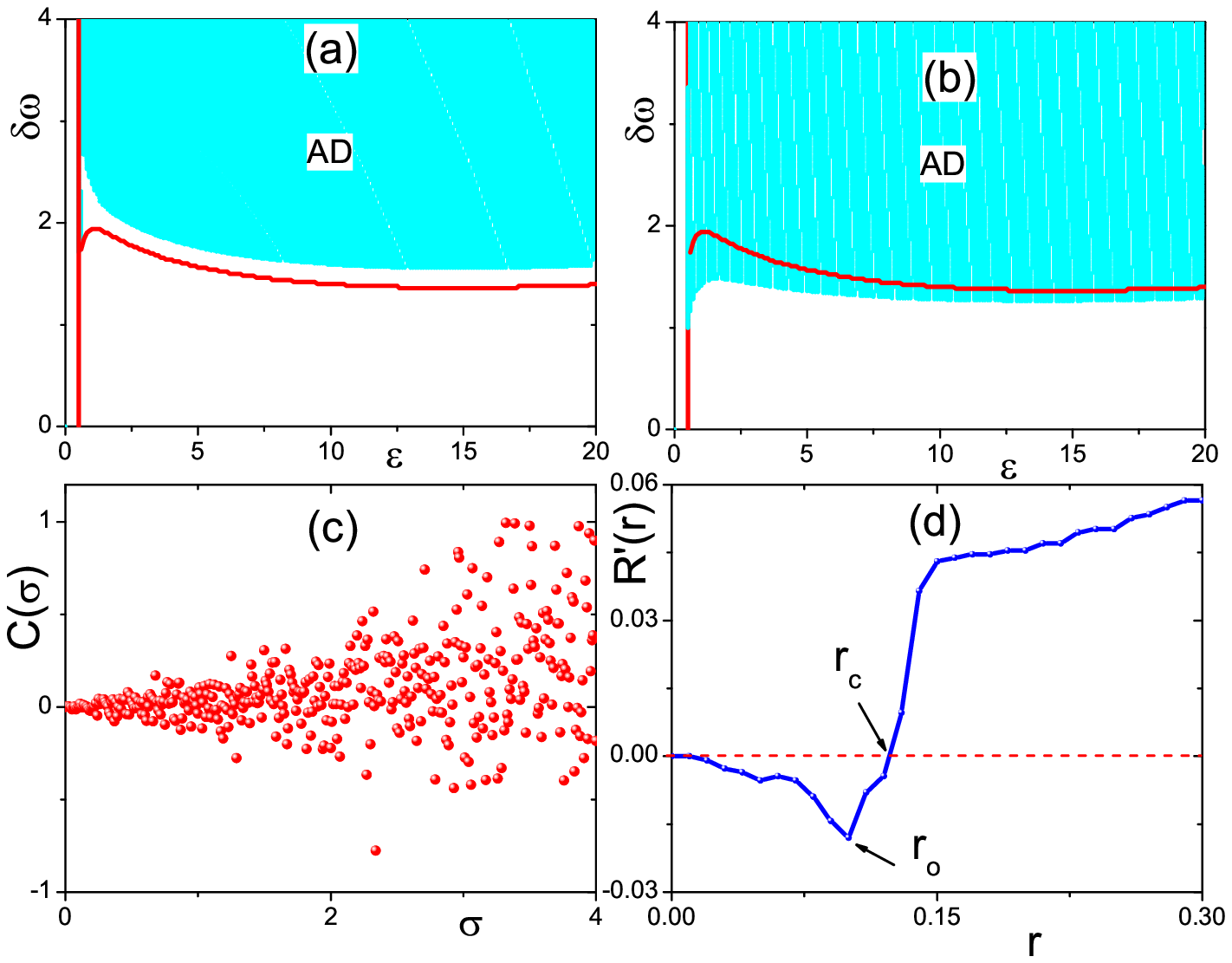}
\caption{(Color online) (a)(b) The effects of noise intensity on AD
domain for gradient coupled oscillators with periodical boundary
conditions($N=10$,$r=5$), AD domain of two realizations of noise
with intensity $\sigma=2$(cyan region). The red curve is the
boundary of AD without noise. The noise can either enlarge or shrink
AD domain. (c)Factors $C(\sigma)=1-S(\sigma)/S(0)$ versus noise
intensity $\sigma$. $S(\sigma)$ is the area of non-AD domain with
noise intensity $\sigma$. The deviation of $C(\sigma)$ is increasing
with deviation of AD domain area. The stronger noise increase the
deviation of AD domain. (d) $R'(r)$ versus r of the gradient coupled
Rossler oscillators with periodical boundary conditions and $N=10$,
where $r_c=0.124, r_o=0.1$.} \label{fig_7}
\end{figure}

It is necessary to make some discussion. For convenience of
analysis, the frequency distribution is set as regular monotonic
trend as in Ref.\cite{rub}. To explore the influence of frequency
distribution on the AD domain, the noise is added to each frequency.
Set $w_j=w_1+(\delta\omega(j-1)+\xi)$, where $\xi$ is the gauss
noise with strength $\sigma$, namely,
$\langle\xi_i\rangle=0$,$\langle\xi_i(t)\xi_j(t')\rangle=\sigma\delta_{ij}\delta(t-t')$.
The AD domain of $N=10$,$\sigma=2$,$r=5$ is presented in
Fig.\ref{fig_7}(a)(b) for different realization of noise. Where the
noise may either enlarge or shrink the AD domain for different
realization of noise. To see the effects of the noise strength
$\sigma$ on AD domain area, the factor $C(\sigma)=1-S(\sigma)/S(0)$
versus $\sigma$ is presented in Fig.\ref{fig_7}(c), where
$S(\sigma)$ is the non-AD domain area for noise constant $\sigma$
for $\epsilon \in(0.5,1.5),\delta\omega\in(0,2)$. Obviously, with
increasing noise intensity, the noise tends to enlarge the deviation
of the factor $C(\sigma)$, the effects of noise on the AD domain
deviation is positively related to the noise intensity $\sigma$.
Secondly, we may point out that the gradient coupling effects on AD
domain is not exclusively existing in the coupled periodical
oscillators but also existing in gradient coupled chaotic
oscillators such as Rossler system
$dx/dt=wy-z,dy/dt=wx+0.165z,dz/dt=0.2+z(x-10)$ with the coupling
scheme as described in Eq.\ref{eq1}, where $w_j=1+(j-1)\delta\omega,
j=1,2,...,N$. The factor $R'(r)$ defined above ($S(r)$ is the non-AD
domain area for corresponding $r$ for $\epsilon
\in(0.5,1.5),\delta\omega\in(0,2)$) versus gradient coupling
constant $r$ for $N=10$ coupled Rossler oscillators with periodical
boundary condition is presented in Fig.\ref{fig_7}(d) where the
increasing $r$ firstly decreases then increases $R'(r)$, which
indicate that the gradient coupling first enlarge then shrink the AD
domain area of coupled Rossler system. Moreover, there is also an
optimal gradient coupling constant $r_o=0.1$ and a critical gradient
coupling constant $r_c=0.124$.

 \textbf{Conclusion}

The effects of gradient coupling effect on AD domain in an array of
coupled nonidentical is strongly related to the boundary conditions.
With no-flux boundary conditions, the gradient coupling tends to
monotonically enlarge the AD domain in the parameter space $\epsilon
\sim \delta \omega$ within the critical gradient coupling constant
$r_c$. When $r>r_c$, the gradient coupling has no effects on AD
domain any more. With the periodical boundary conditions, there is
also a critical gradient coupling constant $r_c$.  When $r>r_c$, the
gradient coupling tends to shrink the AD domain.  When $r\le r_c$,
it first enlarges then decreases the AD domain, that is, there is an
optimal gradient coupling constant $r_o$ to realize largest AD
domain for $N\ge5$. The AD domain of those gradient coupled system
are analytically predicted for small system size number. The
remarkable thing is that the effects of gradient coupling on
frequency mismatches caused AD is quite different with that on the
time-delay induced AD\cite{zou} where gradient coupling
monotonically decreases the AD domain till completely eliminates AD.
Ref.\cite{rub} point out that the frequency mismatch $\delta\omega$
is beneficial to the AD, that is, larger $\delta\omega$ is easier to
become AD in diffusively coupled array of oscillators. In this work,
one may find that the gradient coupling is helpful to AD dynamics
since smaller frequency mismatches $\delta \omega$ is needed for AD
with proper gradient coupling constant. The gradient effects on the
AD in coupled non-identical oscillators would be practically
valuable to the dynamics control.

\textbf{Acknowledgement}

Weiqing Liu is supported by NSFC (Grant Nos 10947117,11062002) and
Science and Technology  Project of Educational Department Jiangxi
Province(Grant No. GJJ10162); Lixiang Li is supported by the
Foundation for the Author of National Excellent Doctoral
Dissertation of PR China (FANEDD) (Grant No. 200951), and
Specialized Research Fund for the Doctoral Program of Higher
Education  (No. 20100005110002).

\newpage

\newpage
\bigskip
\textbf{Reference}
\begin{thebibliography}{99}

\bibitem{win} Winfree A T 1980 The Geometry of Biological Time Springer-Verlag, New York
\bibitem{kur}  Kuramoto Y 1984 Chemical Oscillations, Waves and Turbulence Springer, Berlin
\bibitem{pik} Pikovsky A, Rosenblum M and Kurths J 2001 Synchronization: A Universal Concept in Nonlinear
Dynamics Cambridge University Press, Cambridge, England
\bibitem{bar}  Bar-Eli K 1985 Physica D  \textbf{14 242}
\bibitem{ull} Ullner E,  Zaikin A, Volkov E I and Garc¨ªa-Ojalvo J 2007 Phys. Rev. Lett. \textbf{99
148103}
\bibitem{kos1} Koseska A,Volkov E and Kurths J 2009 Europhys. Lett.  \textbf{85
28002}
\bibitem{kos2} Koseska A, Volkov E and Kurths J 2010 Chaos  \textbf{20
023132}
\bibitem{erm} Ermentrout G B 1990 Physica D  \textbf{41 219-231}
\bibitem{rub} Rubchinsky L and Sushchik M 2000 Phys. Rev. E  \textbf {62
6440}
\bibitem{yang} Yang J Z, 2007 Phys. Rev. E  \textbf{76 016204}
\bibitem{atay} Fatihcan M A, 2003 Physica D  \textbf{183 1¨C18}
\bibitem{hou} Hou Z and Xin H 2003 Phys. Rev. E  \textbf{68 055103R}
\bibitem{liu} Liu W Q, Wang X G , Guan S and Lai C-H 2009 New J. Phys.  \textbf{11
093016}
\bibitem{xiao1} Yang J Z, Hu G and Xiao J H 1998 Phys. Rev. Lett.  \textbf{80
496}
\bibitem{xiao2}Xiao J H, Hu G, Yang J Z and Gao J H 1998 Phys. Rev. Lett.  \textbf{81
5552}
\bibitem{zhan1}Zhan M, Hu G and Yang J Z 2000 Phys. Rev. E  \textbf{62
2963}
\bibitem{zhan2} Zhan M, Gao J H, Wu Y and Xiao J H 2007 Phys. Rev. E  \textbf{76
036203}
\bibitem{zhan3} Zou W and Zhan M 2008 Europhys. Lett.  \textbf{81
10006}
\bibitem{mot} Motter A E, Zhou C S and Kurths J 2005 Europhys. Lett.  \textbf{69 334};  2005 Phys. Rev. E  \textbf{71
016116}
\bibitem{xin1} Wang X G, Liang H,  Lai Y C and Lai C H 2007 Phys. Rev. E  \textbf{76
056113}
\bibitem{xin} Xingang Wang, Cangtao Zhou and Choy Heng Lai, 2008 Phys. Rev. E  \textbf{77
056208}
\bibitem{zou} Zou W, Yao C G and  Zhan M 2010 Phys. Rev. E  \textbf{82
056203}
\bibitem{ram}Ramana Reddy D V, Sen A and Johnston G L 1998 Phys. Rev. Lett.  \textbf{80
5109}
\bibitem{kon} Konishi K 2003 Phys. Rev. E  \textbf{68 067202}
\bibitem{kar} Karnatak R, Ramaswamy R and Prasad A 2007 Phys. Rev. E  \textbf{76
035201R}
\bibitem{zou1} Zou W, Wang X G, Zhao Q and Zhan M 2009 Fron. Phys. China  \textbf{4
97}
\bibitem{wen}  Liu W Y, Xiao J H and Yang J Z 2004 Phys.Rev.E 7 \textbf{0
066211}
\end{thebibliography}
\end{document}